\documentclass[runningheads]{llncs}

\usepackage{epsfig}
\usepackage{amsmath}
\usepackage{amsfonts}

\def\nottoobig#1{{\hbox{$\left#1\vcenter
to1.111\ht\strutbox{}\right.\n@space$}}}

\newcommand{\ie}{$\mbox{i.e.}$}

\newlength{\filength}
\newsavebox{\gcbox}
\sbox{\gcbox}{\framebox[\filength]{\rule{0ex}{2ex}}}

\newcommand{\singlespacing}{\let\CS=
\@currsize\renewcommand{\baselinestretch}{1}\tiny\CS}
\newcommand{\singlespacingplus}{\let\CS=
\@currsize\renewcommand{\baselinestretch}{1.25}\tiny\CS}
\newcommand{\doublespacing}{\let\CS=
\@currsize\renewcommand{\baselinestretch}{1.75}\tiny\CS}
\newcommand{\draftspacing}{\let\CS=
\@currsize\renewcommand{\baselinestretch}{2.0}\tiny\CS}

\def\zo{\{0,1\}}

\usepackage{amsfonts}

\author{Marius Zimand
\thanks{The author is supported in part
by NSF grant CCF 1016158. Most of the results have been presented at International
Workshop on Theoretical Computer Science, Feb 21-24, 2012, Auckland, New
Zealand. URL: \tt  http://triton.towson.edu/\~{ }mzimand}.}
\institute{
{Department of Computer and Information Sciences, Towson University,
Baltimore, MD, USA}
}
\authorrunning{M.~Zimand}

\title{Symmetry of Information: A Closer Look}

\begin{document}

\maketitle

\begin{abstract} Symmetry of information establishes a relation between the
information that $x$ has about $y$ (denoted $I(x : y)$) and the  information
that $y$ has about $x$ (denoted $I(y : x)$). In classical information theory,
the two are exactly equal, but in algorithmical information theory, there is a
small excess quantity of information that differentiates the two terms, caused
by the necessity of packaging information in a way that makes it accessible to
algorithms. It was shown in~\cite{zim:c:symkolm} that in the case of strings with simple complexity (that
is the Kolmogorov complexity of their Kolmogorov complexity is small), the relevant information can be
packed in a very economical way, which leads to a tighter relation between $I(x
: y)$ and $I(y : x)$ than the one provided in the classical
symmetry-of-information  theorem of Kolmogorov and Levin. We give here a simpler proof of this result. This result implies a van Lambalgen-type theorem for finite strings and plain complexity: If $x$ is $c$-random and $y$ is $c$-random relative to $x$, then $xy$ is $O(c)$-random. We show that a similar result holds for prefix-free complexity and weak-K-randomness.
\end{abstract}


\section{Introduction}

In classical information theory, the information that $X$ has about $Y$ is equal to the information that $Y$ has about $X$, \ie, 
\[I(X : Y) = I(Y : X).\footnote{$X$ and $Y$ are random variables and $I(X : Y) = H(Y) - H(Y \mid X)$, where $H$ is Shannon entropy.}
\]

In algorithmical information theory, there exists a similar relation, but it is less perfect:
\[
I(x : y) \approx I(y : x).
\]
In this paper we take a closer look at the level of approximation indicated by ``$\approx$." In the line above, $x$ and $y$ are binary strings of lengths $n_x$ and, respectively, $n_y$, and $I(x : y)$, the information in $x$ about $y$, is defined by $I(x : y) = C(y \mid~n_y) - C(y \mid x, n_y)$, where $C( \cdot)$ is the plain Kolmogorov complexity. The precise form of the  equation is given by the classical Kolmogorov-Levin theorem~\cite{zvo-lev:j:kol}, which states that
\begin{equation}
\label{e:kolmlevin}
|I(x : y) - I(y : x)| = O(\log n_x + \log n_y).
\end{equation}
This is certainly an important result, but in some situations, it is not saying much. Suppose, for example, that $x$ has very little information about $y$, say, $I(x : y)$ is a constant (not depending on $x$ and $y$). What does this tell about $I(y : x)$? Just from the definition and ignoring $I(x : y)$, $I(y : x)$ can be as large as $n_x - O(1)$, and, somewhat surprisingly, given that $I(x : y) = O(1)$, $I(y : x)$ can still have the same order of magnitude. As an example, consider $y$ a $c$-random string (meaning $C(y \mid n_y) \geq n_y - c$), whose length $n_y$ is also $c$-random. Let $x$ be the binary encoding of $n_y$. Then $I(x : y) = C(y \mid n_y) - C(y \mid x, n_y) = O(1)$, but $I(y : x) =  C(x \mid n_x) - C(x \mid y, n_x) = n_x - O(1)$.
This example shows that even though, for some strings, the relation~(\ref{e:kolmlevin}) is trivial,  it is also tight.

In many applications, the excess term $O(\log n_x + \log n_y)$ does not hurt too much. But sometimes it does. For example, it is the reason why the Kolmogorov-Levin Theorem does not answer the following basic ``direct product'' question: If $x$ and $y$ are $c$-random $n$-bit strings and $I(x : y) \leq c$, does it follow that $xy$ is $O(c)$-random? 

The answer is positive  (providing the finite analog of the van Lambalgen theorem). It follows from a recent result of the author~\cite{zim:c:symkolm}, which establishes a more precise symmetry-of-information relation for strings with \emph{simple complexity}, \ie, for strings $x$ such that $C^{(2)}(x \mid n_x)$ is small, say, bounded by a constant.\footnote{The notation $C^{(2)}(x \mid n_x)$ is a shorthand for $C(C(x \mid n_x) \mid n_x)$.} Note that all random strings have simple complexity, and that there are other types of strings with simple complexity as well. The main result in~\cite{zim:c:symkolm} implies that if $x$ and $y$ are strings of equal length that have simple complexity bounded  by $c$ and $I(x : y) \leq c$, then $I(y : x) \leq O(c)$.

The proof method in~\cite{zim:c:symkolm} is based on randomness extraction. 
Alexander Shen (personal communications) has observed that in fact a proof of the above result in~\cite{zim:c:symkolm} can be obtained via the standard method used in the proof of Kolmogorov-Levin Theorem (see~\cite{dow-hir:b:algrandom,li-vit:b:kolmbook}). 
We present here such a proof.


We slightly extend the symmetry-of-information relation from~\cite{zim:c:symkolm} to strings $x$ and $y$ that may have different lengths.  We prove the following (with the convention that $\log 0 = 0$):
\smallskip

\begin{theorem}[Main Theorem] 
\label{t:main}
For all strings $x$ and $y$,
\[
I(y : x) \leq I(x : y) + O(\log I(x : y))  + O(\log |n_x - n_y|) + \delta(x,y),
\]
where $\delta(x,y) = O(C^{(2)}(x \mid n_x) + C^{(2)}(y \mid n_y))$.
\end{theorem}

Thus, for strings $x$ and $y$ with simple complexity, $I(y : x) \leq I(x :~y) + O(\log I(x : y)) + O(\log |n_x - n_y|)$.

As mentioned, Theorem~\ref{t:main} implies an analog of the van Lambalgen Theorem for finite strings and plain Kolmogorov complexity. Recall that van Lambalgen Theorem~\cite{vlam:t:randomseq} states that for infinite sequences $x$ and $y$ the following two statements are equivalent: 

(1) $x$ is random and $y$ is random relative by $x$, 

(2) $x \oplus y$ is random. 

(Here, random means Martin-L\"{o}f random, and $x \oplus y$ is the infinite sequence obtained by alternating the bits of $x$ and $y$).  Theorem~\ref{t:main} implies immediately the following.
\begin{theorem}[\cite{zim:c:symkolm}]
Let $x$ and $y$ be two strings of length $n$ such that $x$ is $c$-random and $y$ is $c$-random relative to $x$. Then $xy$ is $O(c)$-random.

Formally, if $C(x \mid n) \geq n- c, C(y \mid x) \geq n- c$, then $C(xy \mid 2n) \geq 2n - O(c)$.

(The constant in the $O(\cdot)$ notation depends only on the universal Turing machine.)
\end{theorem}

Is there a van Lambalgen Theorem analog for prefix-free complexity? To adress this question, we note that there are two notions of randomness for the prefix-free complexity $K$, weak $K$-randomness, and strong $K$-randomness (see~\cite{dow-hir:b:algrandom}). An $n$-bit string $x$ is 

(1) weakly $c$-(K-random) if $K(x \mid n) \geq n-c$, and 

(2) strongly $c$-(K-random) if $K(x \mid n) \geq n + K(n)-c$. 

For weak $K$-randomness, we prove in Section~\ref{s:prefixfree} the following analog of the van Lambalgen Theorem. For strong $K$-randomness, the question remains open.
\begin{theorem}
\label{t:prefixfree}
Let $x$ and $y$ be two strings of length $n$ such that $x$ is weakly $c$-(K-random) and $y$ is weakly $c$-(K-random) relative to $x$. Then $xy$ is weakly $O(c)$-(K-random).

Formally, if $K(x \mid n) \geq n- c, K(y \mid x) \geq n- c$, then $K(xy \mid 2n) \geq 2n - O(c)$.

(The constant in the $O(\cdot)$ notation depends only on the universal Turing machine.)
\end{theorem}

\subsection{Proof techniques}
The proofs of Symmetry of Information Theorems have a combinatorial nature. To fix some ideas, let us first sketch the proof of the classical Kolmogorov-Levin Theorem which establishes relation~(\ref{e:kolmlevin}). The setting of the theorem, as well as for the rest of our discussion,  is as follows: $x$ and $y$ are binary strings of lengths $n_x$ and, respectively, $n_y$. An easy observation (see Section~\ref{s:symchain}), shows that a relation between $I(y : x)$ and $I(x : y)$ 
 can be deduced from a ``chain rule:" $C(xy \mid n_x, n_y) \geq C(x \mid n_x) + C(y \mid x, n_y) - (\mbox{small term})$.  In our case, to obtain~(\ref{e:kolmlevin}), it is enough to show that 
\begin{equation}
\label{e:sym}
C(xy \mid n_x, n_y) \geq C(x \mid n_x) + C(y \mid x, n_y) - O(\log n_x + \log n_y). 
\end{equation}
Let us consider a $2^{n_x} \times 2^{n_y}$ table with rows indexed by $u \in \zo^{n_x}$ and columns indexed by $v \in \zo^{n_y}$. Let $t = C(xy \mid n_x, n_y)$. We assign boolean values to the cells
of the table as follows. The $(u,v)$-cell in the table is $1$ if $C(uv \mid n_x, n_y) \leq t$ and it is $0$ otherwise. The number of cells equal to $1$ is less than $2^{t+1}$, because there are only $2^{t+1}$ programs of length $\leq t$. Let $m$ be such that the number of $1$'s in the $x$ row is in the interval $(2^{m-1}, 2^m]$. Note that, given $x, n_y$ and $t$, we can enumerate the $1$-cells in the $x$-row and one of them is the $(x,y)$ cell. It follows that
\begin{equation}
\label{e:eq1}
C(y \mid x, n_y) \leq m + O(\log t).
\end{equation}
Now consider the set of rows that have at least $2^{m-1}$ $1$s.  The number of such rows is at most $2^{t+1}/2^{m-1} = 2^{t-m+2}$. We can effectively enumerate these rows if we are given $n_x, n_y, m$ and $t$. Since the $x$ row is one of these rows, it follows that
\begin{equation}
\label{e:eq2}
C(x \mid n_x) \leq t- m + O(\log n_y + \log m + \log t).
\end{equation}
Adding equations~(\ref{e:eq1}) and~(\ref{e:eq2}) and keeping into account that $\log m \leq \log n_y$ and $\log t \leq O( \log n_x + \log n_y)$, the relation~(\ref{e:sym}) follows.

A careful inspection of the proof reveals that the excess term $O(\log n_x + \log n_y)$ is caused by the fact that the enumerations involved in describing $y$ and $x$ need to know $m$ (which is bounded by $C(y \mid n_y)$) and $t = C(xy \mid n_x, n_y)$. In case $x$ and $y$ are $c$-random, it is more economical to use randomness deficiencies. In particular instead of using $t= C(xy \mid n_x, n_y)$, we can do a similar argument based on the randomness deficiency of $xy$, \ie,  $w = (n_x + n_y) - C(xy \mid n_x, n_y)$. This is an observation of Chang, Lyuu, Ti and Shen~\cite{cha-lyu-ti-shen:j:kindep}, who attribute it to folklore. For strings with simple complexity, it is advantageous to express $t$ as $t = C(x \mid  n_x) + C(y \mid x) - w$ because the first two terms have a short description.   This yields a proof of the \emph{Main Theorem} which we present in Section~\ref{s:randomfirst}.

\section{Preliminaries}
\subsection{Notation and background on Kolmogorov complexity}
The Kolmogorov complexity of a string $x$ is the length of the shortest effective description of $x$. There are several versions of this notion. We use here  mainly the \emph{plain complexity}, denoted $C(x)$, and the \emph{conditional plain complexity} of a string $x$ given a string $y$, denoted $C(x \mid y)$, which is the length of the shortest effective description of $x$ given $y$. The formal definitions are as follows.
We work over the binary alphabet $\zo$. A string is an element of $\{0,1\}^*$.
If $x$ is a string, $n_x$ denotes its length.  
Let $M$ be a Turing machine that takes two input strings and outputs one string. For any strings $x$ and $y$, define the \emph{Kolmogorov complexity} of $x$ conditioned by $y$ with respect to $M$, as 
$C_M(x \mid y) = \min \{ |p| \mid M(p,y) = x \}$.
There is a universal Turing machine $U$ with the following property: For every machine $M$ there is a constant $c_M$ such that for all $x$, $C_U(x \mid y) \leq C_M(x \mid~y) + c_M$.
We fix such a universal machine $U$ and dropping the subscript, we write $C(x \mid y)$ instead of $C_U(x \mid y)$. We also write $C(x)$ instead of $C(x \mid \lambda)$ (where $\lambda$ is the empty string).   If $n$ is a natural number, $C(n)$ denotes the Kolmogorov complexity of the binary representation of $n$. We use $C^{(2)}(x | n_x)$ as a shorthand for $C(C(x \mid n_x) \mid n_x)$. For two strings $x$ and $y$, the information in $x$ about $y$ is denoted $I(x : y)$ and is defined as $I(x : y) = C(y \mid n_y) - C(y \mid x, n_y)$.

Prefix-free complexity $K$ is defined in a similar way, the difference being that the universal machine is required to be prefix-free.

In this paper, the constant hidden in the $O(\cdot)$ notation only depends on the universal Turing machine.
\if01
For all $n$ and $k \leq n$, 
$2^{k-O(1)} < |\{x \in \zo^n \mid C(x\mid~n,k) < k\}| < 2^k$.


Strings $x_1, x_2, \ldots, x_k$ can be encoded in a self-delimiting way (\ie, an encoding from which each string can be retrieved) using $n_{x_1} + n_{x_2} + \ldots + n_{x_k} + 2 \log n_{x_1} + \ldots + 2 \log n_{x_k} + O(k)$ bits. For example, $x_1$ and $x_2$ can be encoded as $\overline{(bin (n_{x_1})} 01 \overline{(bin (n_{x_2})} 01 x_1 x_2$, where $bin(n)$ is the binary encoding of the natural number $n$ and, for a string $u = u_1 \ldots u_m$, $\overline{u}$ is the string $u_1 u_1 \ldots u_m u_m$ (\ie, the string $u$ with its bits doubled).
\fi

In this paper, by convention $\log 0 = 0$.

\subsection{Symmetry of Information and the Chain Rule}
\label{s:symchain}
All the forms of the Symmetry of Information that we are aware of have been derived from the \emph{chain rule} and this paper is no exception. The chain rule states that $C(xy \mid n_x, n_y) \approx C(x \mid n_x) + C(y \mid x, n_y)$. For us it is of interest to have an accurate estimation of the ``$\approx$'' relation. It is immediate to see that $C(xy \mid n_x, n_y) \leq C(y \mid n_y) + C(x \mid y, n_x) + 2 C^{(2)}(y \mid n_y) + O(1)$. If we show a form of the converse inequality
\begin{equation}
\label{e:chainrule}
C(xy \mid n_x, n_y) \geq C(x \mid n_x) + C(y \mid x, n_y) - (\mbox{small term}),
\end{equation}
then we deduce that $C(x \mid n_x) - C(x \mid y, n_x) \leq C(y \mid n_y) - C(y \mid x, n_y) + 2 C^{(2)}(y \mid~n_y) + (\mbox{small term})$, \ie,
\[
I(y : x) \leq I(x : y)  + 2 C^{(2)}(y \mid n_y) + (\mbox{small term}).
\]
Thus our efforts will be directed to establishing forms of the relation~(\ref{e:chainrule}) in which (small term) is indeed small.

\section{Proof of the Main Theorem}
\label{s:randomfirst}

We demonstrate Theorem~\ref{t:main}, using the standard proof method of the Kolmogorov-Levin Theorem. 

Let $t_x = C(x \mid n_x), t_y = C(y \mid x, n_y)$ and $t = C(xy \mid n_x, n_y)$. We take $w = t_x + t_y - t$, which is the term called \emph{(small term)} in equation~(\ref{e:chainrule}), and plays the role of randomness deficiency. Our goal is to show that 
\[
w = O(\log I(x : y)) + O(\log |n_x - n_y|) + \delta(x,y),
\]
from which the Main Theorem follows (using the discussion in Section~\ref{s:symchain}).

We assume that $w > 0$, otherwise there is nothing to prove. We will need to handle information 
$(|n_x-n_y|, t_x, t_y, w, b))$, where $b$ is a bit that indicates if $n_x > n_y$ or not. Let $\Lambda$ be a string encoding in a self-delimiting way this information, and let $\lambda$ be the length of $\Lambda$. Note that 
\[
\lambda \leq 2 \log w + O(C^{(2)}(x \mid n_x) + C^{(2)}(y \mid n_y) + \log I(x : y) + \log |n_x - n_y|).
\]  
We build a $2^{n_x} \times 2^{n_y}$ boolean table, with rows and columns indexed by the strings in $\zo^{n_x}$ and, respectively, $\zo^{n_y}$, and we set the value of cell $(u,v)$ to be $1$ if $C(uv \mid n_x, n_y) \leq t$, and to be $0$ otherwise. Let $S$ be the set of $1$-cells, and $S_u$ be the set of $1$-cells in row $u$. Note that
\[
|S| \leq 2^{t+1}.
\]
Let $m$ be defined as $2^{m-1} < |S_x| \leq 2^{m}$. We take $F$ to be the set of ``fat" rows, \ie, the set of rows having more than $2^{m-1}$ $1$-cells.  We have $|F| < |S|/2^{m-1} \leq 2^{t-m+2}$. Note that $x$ is in $F$, and that the elements of $F$ can be effectively enumerated given the information $\Lambda$ and $m$. It follows that, given $\Lambda$ and $m$, the string $x$ can be described by its index in the enumeration of $F$. This index can be written on \emph{exactly} $t-m+2$ bits, so that knowing $t$ which can be deduced from $\Lambda$, we can reconstruct $m$. It follows that
\begin{equation}
\label{e:tmtwo}
C(x \mid n_x, \Lambda) \leq t-m+2 + O(1).
\end{equation}
Next we note that $y$ is in $S_x$ and the elements of $S_x$ can be enumerated given $x$ and $\Lambda$. It follows that $y$ can be described by its index in the enumeration of $S_x$. We obtain
\begin{equation}
\label{e:ym}
C(y \mid  x, \Lambda) \leq m + O(1).
\end{equation}
From Equations~(\ref{e:tmtwo}) and~(\ref{e:ym}), we conclude that
\[
C(x| n_x, \Lambda) + C(y| x, \Lambda) \leq t+ O(1) = t_x + t_y - w + O(1).
\]
Note that $t_x = C(x \mid n_x) \leq C(x \mid n_x, \Lambda) + O(\lambda$) and
$t_y = C(y \mid x, n_y) \leq C(y \mid x, \Lambda) + O(\lambda)$.
We obtain $t_x + t_y \leq t_x + t_y - w + O( \lambda)$. Taking into account the estimation for $\lambda$, we have $w - O(\log w) = O(C^{(2)}(x \mid n_x) + C^{(2)}(y \mid n_y) + \log I(x : y) + \log |n_x - n_y|)$, from which, $w = O(C^{(2)}(x \mid n_x) + C^{(2)}(y \mid n_y) + \log I(x : y) + \log |n_x - n_y|)$, as desired.

\section{A van Lambalgen Theorem for prefix-free complexity}
\label{s:prefixfree}

In this section we prove Theorem~\ref{t:prefixfree}.

We introduce first the following notation: $\overline{d}$ is a self delimited encoding of the integer $d$. We can have $|\overline{d}| \leq 2 \log d +4$.

For a string $u$, let $A_u(d) = \{v \in \zo^n \mid K(uv \mid n) \leq 2n-d\}$, and let 
$F(d) = \{u \in \zo^n \mid |A_u(d)| \geq 2^{n-d}\}$.
\smallskip

Consider the following program $p_1$, which has $x$ of length $n$ on the oracle tape (as conditional information):
\smallskip

On input the pair of positive integers $(d, i)$, encoded as $\overline{d} ~\mbox{bin}(i)$, check if $i$ is written on $n - d$ bits. If not, diverge. If yes, enumerate strings $u$ of length $n$ such that $K(xu \mid n) \leq  2n - 2d$. Output the $i$-th string in the enumeration.

Note that the domain of $p_1$ is prefix-free. Let $c_1$ be the length of program $p_1$.
\medskip

Let $p_2$ be the following program, which has $n$ on the oracle tape (as conditional information):
\smallskip

On input the pair of positive integers $(d, i)$, encoded as $\overline{d} ~\mbox{bin}(i)$,  check if $i$ is written on $n - d$ bits. If not, diverge. If yes, enumerate the elements of $F(d)$ and output the $i$-th one.

Program $p_2$ is prefix-free. Let $c_2$ be its length.
\medskip

Let $d_0 = \max \{2(c+4+c_1), 2(c+4+c_2)\}$.
\medskip

If there is some $i$ such that $p_1(d_0,i)$ outputs $u$, then $K(u \mid x) \leq n-d_0 + 2 \log d_0 + 4 + c_1 < n-c$. So, there is no $i$ such that $p_1(d_0, i)$ outputs $y$.  This can happen only if either

(a) $K(xy \mid n) > 2n - 2d_0$, or

(b) $K(xy \mid n) \leq 2n- 2d_0$ but there are $2^{n-d_0}$ strings $u$ which are enumerated before $y$. 

If (a), we are done.

We show that (b) is impossible.

Indeed, if (b) holds, then $|A_x(d_0)| \geq 2^{n-d_0}$ and thus $x$ is in $F(d_0)$. Also $|F(d_0)| \leq 2^{2n-2d_0}/2^{n-d_0} = 2^{n-d_0}$.  It follows that for some $i$, program $p_2$ on input $(d_0, i)$ outputs $x$.  But then $K(x \mid n) < n - d_0 + 2\log d_0 + 4 + c_2 < n - c$, which is a contradiction.

\section{Final remarks}
For symmetry of information, the argument from~\cite{zim:c:symkolm} based on randomness extractors, a concept introduced and studied intensively in computational complexity, can be replaced by the simpler and more direct argument suggested by Sasha Shen and presented here. Nevertheless, we still think that extractors play a role in Kolmogorov complexity theory. They have been used in Kolmogorov complexity extraction (see~\cite{fhpvw:j:extractKol,hit-pav-vin:j:Kolmextraction,mus:t:spaceboundedextractor,zim:j:extractKolm,zim:c:kolmlimindep}, and the survey paper~\cite{zim:j:kolmextractsurvey}) and, of course, this is not surprising given the analogy between Kolmogorov complexity extraction and randomness extraction. But they have also been used for obtaining results that are not immediately connected to randomness extraction, such as counting results regarding dependent and independent strings~\cite{zim:c:countingstrings}, and independence amplification~\cite{zim:c:impossibamplific}.

\section{Acknowledgements} I am grateful to Alexander Shen who has noticed that the main theorem can be obtained with the method used in the standard proof of the Kolmogorov-Levin Theorem.

\bibliography{c:/book-text/theory}
\bibliographystyle{alpha}

\if01
\newcommand{\etalchar}[1]{$^{#1}$}

\fi
\end{document}